\documentclass[aps,pre,amsmath,amsfonts,amssymb,floatfix,superscriptaddress,nofootinbib]{revtex4}
\usepackage{mathrsfs} \usepackage{graphicx,color} \usepackage{bbm} \usepackage{multirow}

\let\a=\alpha \let\b=\beta \let\g=\gamma \let\d=\delta
\let\e=\varepsilon \let\z=\zeta  \let\k=\kappa
 \let\m=\mu   
\let\s=\sigma \let\t=\tau \let\f=\varphi 
   
\let\D=\Delta \let\Th=\Theta  
   
\let\ee=\epsilon \let\r=\rho \let\th=\theta \let\io=\infty

\def\FF{{\cal F}}

\def\DD{{\cal D}}

\def\olx{{\overline{x}}} 
\def\oly{{\overline{y}}} 

\def\ol{\overline}

\def\to{\rightarrow}

\newcommand{\beq}{\begin{equation}} \newcommand{\eeq}{\end{equation}}
\newcommand{\wh}{\widehat} \newcommand{\wt}{\widetilde}

\allowdisplaybreaks

\begin{document}

\title{
The shear modulus of glasses: \\
results from the full
replica symmetry breaking solution
} 

\author{Hajime Yoshino}
\affiliation{Department of Earth and Space Science, Faculty of Science,
 Osaka University, Toyonaka 560-0043, Japan}

\author{Francesco Zamponi}
\affiliation{LPT,
\'Ecole Normale Sup\'erieure, UMR 8549 CNRS, 24 Rue Lhomond, 75005 Paris, France}

\begin{abstract}
We compute the shear modulus of amorphous hard and soft spheres, 
using the exact solution in infinite spatial dimensions that has been developed recently.
We characterize the behavior of this observable in the whole
phase diagram, and in particular around the glass and jamming transitions. Our results are consistent with other theoretical approaches,
that are unified within this general picture, and they are also consistent with numerical and experimental results.
Furthermore, we discuss some properties of the out-of-equilibrium dynamics after a deep quench close to the jamming transition, and
we show that a combined measure of the shear modulus and of the mean square displacement allows one to probe experimentally the complex structure
of phase space predicted by the full replica symmetry breaking solution.
\end{abstract}

\maketitle


\section{Introduction}

Amorphous soft spheres with repulsive interaction, 
that are a good model for colloidal glasses, molecular glasses, and granular media~\cite{TS10,PZ10,BB11}, 
display a very complex rheological behavior, controlled by two distinct ``critical'' points: the glass transition and the jamming transition~\cite{IBS12,PZ10}.
The control parameters for these systems are density (or better packing fraction $\f$) and temperature $T$ (or better
the ratio $T/\ee$ between temperature $T$ and the typical interaction energy $\ee$).
At densities below the glass transition, the system is liquid, hence it can flow and its shear modulus vanishes.
At the glass transition, the system acquires a finite rigidity through 
a discontinuous transition at which the shear modulus jumps 
to a finite value~\cite{SF11,Yo12}.
Simultaneously the yield stress may also
exhibit a similar discontinuous transition~\cite{IBS12,FC02}.
It is important to stress that, at densities slightly above the glass transition and at low enough temperatures, 
rigidity has an {\it entropic} origin, as in a hard sphere crystal~\cite{mason1995linear,IBS12}.
In fact, at zero temperatures, the spheres
have enough room to fit in space without touching, hence they do not interact and the system is loose. At small finite temperature, however, the
spheres vibrate and collide, giving rise to an effective interaction that makes the system rigid~\cite{BW06}.
In this region of density, and in the limit of vanishing temperature,
both the pressure and the shear modulus are therefore proportional to temperature. 
Upon further compression, the system reaches a ``random close packing'' or ``jamming'' density~\cite{PV86,BM60,LNSW10,OLLN02}.
We denote $\f_j$ the packing fraction at the jamming point.
Above this density, the spheres cannot fit in 
space without being ``squeezed'' by their neighbors, and rigidity and pressure acquire a {\it mechanical} origin, being
due to a direct interaction between the particles~\cite{Mason97,IBS12}. 
Both pressure and the shear modulus, therefore, have a finite
zero-temperature limit in the region $\f > \f_j$.
Note that, as a consequence, infinitely hard particles cannot be compressed beyond the jamming point, because the mechanical repulsion becomes infinite
and cannot be overcome by any finite pressure. Hard spheres can therefore be thought as a particular case of soft spheres with repulsive strength $\ee$, 
that is obtained by taking the limit $T/\ee \to 0$ for $\f < \f_j$. Obviously the limit $T/\ee \to 0$ can be taken equivalently by sending $T\to 0$ or
$\ee \to\io$~\cite{IBS12,IBB12,BW09,BJZ11}.

The interplay of the jamming and glass transitions gives rise to complex flow curves in the 
packing fraction-temperature ($\f$, $T$) plane, which have been the subject of many experimental,
numerical and analytical investigations.
A very complete characterization of the flow curves in the regime of interest for the present work has been reported in~\cite{IBS12}. 
Concerning the behavior of the shear modulus $\mu$, most of the previous studies agree on two main qualitative observations.
First, the shear modulus
jumps discontinuously at the glass transition~\cite{IBS12,Yo12,Yo12b,FC02,SF11}; 
secondly, it has a critical behavior around the jamming transition at $\f_j$~\cite{BW06,OLLN02,ZS11,Wyart,DLBW14}.
In fact, in the hard spheres regime
when $T\to 0$ and $\f\to\f_j^-$,
the shear modulus is proportional to $T$ and $\mu/T \sim (\f_j - \f)^{-\k}$,
with $\k$ close to $3/2$~\cite{BW06}.
For soft harmonic spheres at $T=0$, when $\f\to\f_j^+$
the shear modulus vanishes as $\mu \sim (\f - \f_j)^{\k'}$  
with $\k'$ close to $1/2$~\cite{OLLN02,Wyart}.
The exponent $\k$ has been related to other exponents that characterize criticality at the jamming transition~\cite{DLBW14}, 
whereas for soft spheres the prediction that $\k'=1/2$ was made using different approximations~\cite{Wyart,DLDLW14}.
Remarkably, despite the fact that both the glass and the jamming transition happen out-of-equilibrium, hence at protocol-dependent densities, 
the critical properties around these transitions are universal and independent
of the protocol, for a wide range of reasonable preparation protocols~\cite{PZ10,IBS12,CCPZ12,LDW13,DLBW14}.

Constructing a first principle theory able to describe the complex rheological properties of these systems is a difficult task.
One of the most successful theories is Mode-Coupling Theory (MCT)~\cite{Go09}, that is based on an approximate set of dynamical equations.
MCT predicts the existence of sharp dynamical glass transition on a line $T_{\rm d}(\f)$ that ends, for $T\to 0$, 
at density $\f_{\rm d} < \f_j$~\cite{BW09}. For $T<T_{\rm d}(\f)$ (or $\f > \f_{\rm d}$ in the hard sphere limit $T/\ee \to 0$), diffusion is completely arrested and particles
are completely caged by their neighbors.
MCT can describe well the rheological properties
around the glass transition~\cite{FC02,SF11}. However, MCT does not provide good results deep in the glass phase and in particular
it fails to describe correctly the jamming transition~\cite{IB13}.
Another first-principle approach to glass physics is based on the assumption that glasses are long-lived metastable states, and can be
described by a restricted equilibrium thermodynamics. Concrete computations can then be done using replicas, and have been usually
done within the so-called ``1-step replica symmetry breaking'' (1RSB) scheme~\cite{MP09,PZ10}.
This method has been applied to describe the 
rigidity of structural glasses, including hard and soft sphere systems~\cite{YM10,Yo12,Yo12b,OY13}, and provides good qualitative
and quantitative results for the shear modulus, but in the simplest 1RSB scheme it fails to predict the exponents $\k$ and $\k'$ correctly~\cite{Yo12b,OY13}.
Both the MCT and the replica approach are thought to be part of the more general Random First Order Transition (RFOT) scenario for the glass
transition~\cite{KW87,KT87,KTW89}.

Following a well-established tradition in theoretical physics~\cite{1N}, in~\cite{KPZ12,KPUZ13,CKPUZ13,nature}
a new approach was adopted by solving exactly the hard sphere model 
in the limit of $d\to\io$, in which the RFOT scenario is exactly realized~\cite{KW87}.
In particular, in~\cite{CKPUZ13,nature} it was shown that in addition to the glass and jamming transitions, a new transition takes place
deep in the glass phase and before jamming occurs. This so-called Gardner transition is a transition where the 1RSB structure changes
into a full replica symmetry breaking (fullRSB) one~\cite{Ga85,MPV87}. Physically, this corresponds to a splitting of glass basins into
a complex hierarchy of subbasins, akin to the one of the Sherrington-Kirkpatrick model. This structure has been described in detail in
the literature~\cite{MPV87}. In~\cite{CKPUZ13,nature} it was shown that this exact solution predicts a phase diagram which is in very good
qualitative agreement with the one observed in numerical simulations and experiments, and in particular it gives correct {\it analytical} predictions for
the critical exponents that characterize the jamming transition.
The solution, originally obtained for hard spheres,
can be also extended to soft spheres at low enough temperatures in the vicinity of the jamming transition~\cite{BJZ11,CKPUZ13}.

The aim of this paper is to extend the analysis of~\cite{CKPUZ13} to describe the rheological properties of soft and hard spheres in the limit $d\to\io$,
and in particular to compute the shear modulus. Our main results are: {\it (i)} at the dynamic glass transition $\f_{\rm d}$, 
the shear modulus $\mu$ jumps discontinuously
to a finite value, and at densities slightly above the transition it displays a square-root singularity $\mu \sim \mu_{\rm d} + C (\f - \f_{\rm d})^{1/2}$; 
{\it (ii)} the critical properties around the jamming transition
$\f_j$ are the ones described above, with $\mu \sim T (\f_j - \f)^{-\k}$ for $\f\to\f_j^-$ with $\k = 1.41575$, while
the exponent $\k'$ remains for the moment undetermined;
{\it (iii)} we derive predictions for the out-of-equilibrium dynamics after a deep quench close to the jamming transition that
are direct consequences of the fullRSB structure.
Note that, as discussed in~\cite{CKPUZ13}, non-trivial exponents emerge from the mean field computation due to the complex
pattern of fullRSB.
Our approach unifies, in a well controlled mean field setting, several theoretical approaches such as MCT~\cite{FC02,SF11,IB13} 
and effective medium approaches~\cite{Wyart,BW06,DLBW14}.

Whether these infinite-dimensional results can be immediately translated to experimental systems in $d=2,3$ is of course a crucial question
that remains open from the purely theoretical point of view. However,
the qualitative phase diagram is consistent with numerical results~\cite{IBS12}, and the predicted critical exponents also agree well with numerical 
results~\cite{OLLN02,BW06,IBS12,OY13}. Moreover,
the method of solution that is exact in infinite dimensions can be used, in an approximate way, 
to provide quantitative predictions for non-universal
quantities (such as the transition densities $\f_{\rm d}$ and $\f_j$, the equation of state of the glass, the correlation function, etc.) 
in finite dimensions~\cite{MP09,PZ10,BJZ11,Yo12}, which agree well with numerical data.
Despite all these positive results, several problems remain open and we discuss them in the conclusions.

\section{Shearing the molecular liquid}
\label{sec:cloned}

\subsection{General formulation}

In the following, we consider a system of $N$ identical particles in a cubic $d$ dimensional volume $V$, with
density $\r = N/V$.
Particles interact through a two-body potential $v(r)$, which in most cases will be the hard sphere
potential with diameter $D$. 
To keep the discussion more general, we consider also a generic soft
potential with range $D$ and temperature $T = 1/\b$, defined as
$v(r) = \ee (1-r/D)^2 \th(D-r)$,
where $\th(x)$ is the Heaviside step function. The hard sphere limit correspond to $\ee\to\io$.
We define the packing fraction $\f = \r V_d (D/2)^d$, where $V_d$ is the volume of a $d$-dimensional
sphere of unit radius~\cite{PZ10}, and a scaled packing fraction $\wh\f = \r V_d D^d/d = 2^d \f/d$.
We consider first the thermodynamic limit $N\to\io$, and then the limit $d\to\io$,
which is natural from the statistical mechanics point of view. Also, the inverse limit is ill-defined because
a minimal number of particles $N \sim 2^d$ is needed to define properly the problem in dimension $d$~\cite{CIPZ11}.

A general approach to compute properties of glasses through a ``cloned liquid'' replica method
has been presented in~\cite{MP09}, and applied to the computation of the shear modulus in~\cite{YM10,Yo12}.
To keep this paper reasonably short, we cannot reproduce this construction here.
We therefore assume that the reader is familiar with
{\it (i)} the general formalism of the replica method of spin glasses, including the 1-step
replica symmetry breaking (1RSB) and full replica symmetry breaking (fullRSB) schemes, reviewed e.g.
in~\cite{CC05,MPV87};
{\it (ii)} the ``cloned liquid'' construction to compute properties of glasses,
originally introduced in~\cite{Mo95,MP96} and reviewed in~\cite{MP09}, including its application to compute
the shear modulus~\cite{Yo12};
{\it (iii)} its application to hard sphere systems, reviewed in~\cite{PZ10}, and in particular
the structure of the 1RSB solution in the limit $d\to\io$~\cite[Sec.VI]{PZ10};
{\it (iv)} the construction of a fullRSB solution for hard spheres in $d\to\io$ and the associated
phase diagram, obtained in~\cite{KPZ12,KPUZ13,CKPUZ13}.

To describe glassy states,
we consider $m$ identical replicas of the original system~\cite{Mo95}.
To describe the fullRSB structure, we follow the results and the notations of~\cite{CKPUZ13}, 
and we assume that the $m$ replicas all belong to one of the 
largest metabasins, in such a way that their mean square displacement is at most $\wh\D_1$.
As a consequence, the replicas form a {\it molecular liquid} in which each
molecule contains one particle of each of the replicas.

Each molecule is described by a set of $m$ coordinates $\olx = \{ x_1 \cdots x_m\}$, where $x_a$ are $d$-dimensional vectors.
Following~\cite{YM10,Yo12}, we apply a shear-strain $\g_a$ to replica $a$, which is obtained by deforming linearly
the volume in which the system is contained. 
We call $x'_\m$, with $\mu=1,\cdots,d$, the coordinates in the original reference frame, in which the shear-strain is applied.
In this frame, the cubic volume is deformed because of shear-strain.
To remove this undesirable feature, we introduce new coordinates $x_\m$ of a ``strained'' frame in which the volume
is brought back to a cubic shape.
If the shear-strain is applied along direction $\mu=2$, then
for replica $a$ all the coordinates are unchanged, $x_{a\mu} = x_{a\mu}'$, except the first
one which is changed according to
\beq
x_{a1}' = x_{a1} + \g_a x_{a2} \ , \hskip30pt  x_{a1} = x_{a1}' - \g_a x_{a2}' \ .
\eeq
Let us call $S(\g_a)$ the matrix such that $x_a' = S(\g_a) x_a$.
In the original frame (where the volume is deformed by strain), two particles of replica $a$
interact with the potential $v(|x_a'-y_a'|)$.
If we change variable to the strained frame (where the volume is not deformed), the interaction is 
\beq
v_{\g_a}(x_a-y_a) = v(|S(\g_a)(x_a-y_a)|) \ . 
\eeq
An important remark is that $\det S(\gamma) =1$ meaning 
that the simple shear-strain defined above does not
change the volume and thus the average density $\rho=N/V$ of the system.

In $d\to\io$, we keep only the first non-trivial term of the virial series~\cite{FP99,PZ10,KPZ12}.
Using the coordinates of the strained frame, the system is translationally invariant in the usual way
and, following the same steps and notations of~\cite{KPZ12},
we can write the free energy of the replicated
liquid as
\beq\label{eq:free}
\begin{split}
-\b N F(\{\g_a\}) & = \int d\overline{x} \r(\overline{x}) [1 - \log \r(\overline{x}) ] + \frac12 \int d\overline{x} d\overline{y} 
\r(\overline{x})\r(\overline{y})f_{\{\g_a\}}(\overline{x}-\overline{y}) \\
& = V \int \DD\overline{u} \r(\overline{u}) [1 - \log \r(\overline{u}) ] + \frac{V}2 \int \DD\overline{u} 
\DD\overline{v} 
\r(\overline{u})\r(\overline{v}) \ol f_{\{\g_a\}}(\overline{u}-\overline{v}) \ ,
\end{split}
\eeq
where $X = m^{-1} \sum_a x_a$ is the center of mass of a molecule, $u_a = x_a -X$ is the displacement of
replica $a$ in the molecule,
the Mayer function is
\beq\label{eq:Mayer}
f_{\{\g_a\}}(\overline{x}-\overline{y}) = -1 + \prod_{a=1}^m e^{-\b v(|S(\g_a)(x_a-y_a)|)} = f( \{ S(\g_a)(x_a-y_a) \} ) \ ,
\eeq
$f(\overline{x}-\overline{y})$ is the usual Mayer function that was considered in~\cite{KPZ12},
and $\ol f_{\{\g_a\}}(\overline{u}-\overline{v}) = \int dX f_{\{\g_a\}}(X + \overline{u}-\overline{v})$ 
is the translationally
averaged Mayer function as in~\cite{KPZ12}.

Following~\cite{YM10,Yo12}, we want to develop the free energy for small $\g_a$, and we define
\beq\label{eq:Fexp}
F(\{\g_a\}) =  F(\{ 0\}) + \sum_{a=1}^m \s_a \g_a + \frac{1}2 \sum_{a,b}^{1,m} \mu_{ab} \g_a\g_b + \cdots
\eeq
which defines the shear modulus matrix $\mu_{ab}$, whose physical interpretation will be discussed later.

\subsection{Replicated Mayer function in presence of shear-strain}

Clearly, the shear-strain $\g_a$ enters only in the second term of Eq.~\eqref{eq:free}, that represents
the mean field density-density interaction, and for this reason we call it 
the ``interaction term''~\cite{KPZ12}.
We shall therefore repeat the steps of~\cite[Sec.~5]{KPZ12} in presence of the shear-strain.

We specialize here to hard spheres for simplicity, but the derivation can be easily extended to soft spheres
following the analysis of Ref.~\cite{CKPUZ13}.
Taking into account translational invariance following the same
steps as in~\cite{KPZ12}, we obtain a modified Mayer function that reads
\beq
\ol f_{\{\g_a\}}(\bar u ) = \int dX \left\{  -1 + \prod_{a=1}^m \theta(|S(\g_a) (X + u_a) | - D) \right\}
= - \int dX \,  \theta( D - \min_a | S(\g_a)( X + u_a ) | ) \ ,
\eeq
where $\theta(x)$ is the Heaviside step function.
This function is integrated over $\bar u - \bar v$ in Eq.~\eqref{eq:free}, where $\bar u$ and $\bar v$
are extracted from the density distribution $\r(\bar u)$. Here we shift $u_a - v_a \to u_a$.
The $m$ vectors $u_a$ define an $m$-dimensional plane in the $d$-dimensional space,
and because the vectors $u_a$ are extracted at random, this plane is orthogonal to the shear-strain
directions $\mu=1,2$ with probability going to 1 for $d\to\io \gg m$.
Hence, the vector $X$ can be decomposed in a two dimensional
vector $\{X_1,X_2\}$ parallel to the shear-strain plane, 
a $(d-m-2)$-component
vector $X_\perp$, orthogonal to the plane $\mu=1,2$ and to the plane defined by $u_a$,
and a $m$-component vector $X_\parallel$ parallel to that plane.
Defining $\Omega_d$ as the $d$-dimensional solid angle and recalling that $V_d=\Omega_d/d$,
and following the same steps as in~\cite[Sec.~5]{KPZ12},
we have
\beq\begin{split}
\ol f_{\{\g_a\}}(\bar u ) & 
=  - \int dX \,  \theta( D^2 - \min_a |S(\g_a) ( X + u_a) |^2 ) \\
&= -\int dX_1 dX_2 \ d^{m}X_\parallel \ d^{d-m-2}X_\perp \ 
\theta( D^2 - \min_a\{  (X_1 + \g_a X_2 )^2 + X_2^2 + |X_\parallel + u_a |^2 + | X_\perp |^2 \} ) \\
&= - \Omega_{d-m-2} \int  dX_1 dX_2 \ d^{m}X_\parallel \int_0^\io dx \, x^{d-m-3} \,  
\theta( D^2 - x^2 - \min_a\{ (X_1 + \g_a X_2 )^2 + X_2^2 + | X_\parallel + u_a |^2 \}) \\
&= - \Omega_{d-m-2} \int  dX_1 dX_2 \ d^{m}X_\parallel \int_0^{ \sqrt{D^2 - \min_a\{ (X_1 + \g_a X_2 )^2 + X_2^2 + | X_\parallel + u_a |^2 \}} }  dx \, x^{d-m-3} \\
& = - V_{d-m-2}  \int  dX_1 dX_2 \ d^{m}X_\parallel \, \Theta_{d-m-2}(D^2  - \min_a\{ (X_1 + \g_a X_2 )^2 + X_2^2 + | X_\parallel + u_a |^2 \} )
\end{split}\eeq
where we defined the function~\cite{KPZ12}
\beq
\Th_{n}(x) = x^{n/2} \th(x) \ .
\eeq

It has been shown in~\cite{KPZ12} that
the region where $\overline{f}_{\{\g_a\}}(\bar u)$ has a non-trivial dependence on the $u_a$ is 
where $u_a \sim 1/\sqrt{d}$.
Hence we define
$u_a = x_a D/\sqrt{d}$, $X_{1,2} = \z_{1,2} D/\sqrt{d}$
and $X_\parallel = \ee D/\sqrt{d}$. 
Using that $\lim_{n\to\io} \Th_n(1+y/n) = e^{y/2}$, 
and that for large $d$ and finite $k$ we have $V_{d-k}/V_d \sim d^{k/2} / (2\pi)^{k/2}$,
we have
\beq\begin{split}
\ol f_{\{\g_a\}}(\bar u ) 
& = - \frac{V_{d-m-2}}{V_d} \frac{V_d D^{d}}{d^{(m+2)/2}} \int  d\z_1 d\z_2  d^{m}\ee \, 
\Theta_{d-m-2}\left(1  - \frac1d \min_a\{ (\z_1 + \g_a \z_2 )^2 + \z_2^2 + | \ee + x_a |^2 \} \right) \\
& \sim - V_d D^d  \int \frac{ d\z_1 d\z_2  d^{m}\ee }{(2\pi)^{(m+2)/2}} 
\, e^{  - \frac12 \min_a \{ (\z_1 + \g_a \z_2 )^2 + \z_2^2 + | \ee + x_a |^2 \}  } 
\equiv - V_d D^d \FF_{\{\g_a\}}(\bar x )
\ ,
\end{split}\eeq
where the function $\FF$ has been introduced following~\cite{KPZ12,KPUZ13}.

We can then follow the same steps as in~\cite[Sec.V C]{KPUZ13} to write
\beq
\begin{split}
\mathcal F_{\{\g_a\}}(\bar x) &= 
\int \frac{ d\z_1 d\z_2  d^{m}\ee }{(2\pi)^{(m+2)/2}} \, e^{  - \frac12 \min_a \{ (\z_1 + \g_a \z_2 )^2 + \z_2^2 + | \ee + x_a |^2 \}  } =
\int \frac{ d\z_1 d\z_2  d^{m}\ee }{(2\pi)^{(m+2)/2}}
\lim_{n\to 0} \left( \sum_{a=1}^m e^{-\frac1{2n} [ (\z_1 + \g_a \z_2 )^2 + \z_2^2 + | \ee + x_a |^2 ] }
\right)^n
\\
&=\lim_{n\to 0}\sum_{n_1,\ldots, n_m; \sum_a^mn_a=n}\frac{n!}{n_1!\ldots n_m!}
\int \frac{ d\z_1 d\z_2  d^{m}\ee }{(2\pi)^{(m+2)/2}}
e^{ - \sum_a \frac{n_a}{2n} [ (\z_1 + \g_a \z_2 )^2 + \z_2^2 + | \ee + x_a |^2 ]     } \\
\\
&=\lim_{n\to 0}\sum_{n_1,\ldots, n_m; \sum_a^mn_a=n}\frac{n!}{n_1!\ldots n_m!}e^{ -\frac 12 \sum_{a=1}^m \frac{n_a}{n}|x_a|^2+\frac{1}{2} \sum_{a,b}^{1,m}\frac{n_an_b}{n^2}x_a\cdot x_b } 
\int \frac{ d\z_1 d\z_2 }{2\pi}
e^{ - \sum_a \frac{n_a}{2n} [ (\z_1 + \g_a \z_2 )^2 + \z_2^2  ]     }
\\
&=\lim_{n\to 0}\sum_{n_1,\ldots, n_m; \sum_a^mn_a=n}\frac{n!}{n_1!\ldots n_m!}e^{- \frac{1}{4} \sum_{a,b}^{1,m}\frac{n_an_b}{n^2} (x_a - x_b)^2 } 
\int \frac{ d\z }{\sqrt{2\pi}}
e^{ - \frac{\z^2}2 \left[ 1 + \frac12 \sum_{ab} \frac{n_a n_b}{n^2} (\g_a - \g_b)^2 \right]    }
\\
&=\lim_{n\to 0}\sum_{n_1,\ldots, n_m; \sum_a^mn_a=n}\frac{n!}{n_1!\ldots n_m!}e^{- \frac{1}{4} \sum_{a,b}^{1,m}\frac{n_an_b}{n^2} \D_{ab} } 
\int \frac{ d\z }{\sqrt{2\pi}}
e^{ - \frac{\z^2}2 \left[ 1 + \frac12 \sum_{ab} \frac{n_a n_b}{n^2} (\g_a - \g_b)^2 \right]    } \ ,
\end{split}\eeq
where we introduced the matrix $\hat\D$ of mean square displacements between replicas
\beq\label{eq:Dabdef}
\D_{ab} = (x_a - x_b)^2 = \frac{d}{D^2} (u_a - u_b)^2 \ .
\eeq
This quantity encodes the mean square displacements in, and between, glassy states, as discussed in~\cite{PZ10,KPZ12,CKPUZ13}. We come back on this point
in the following.
We should now recall that in Eq.~\eqref{eq:free} the Mayer function is evaluated in $\bar u - \bar v$,
hence after rescaling the function $\FF$ is evaluated in $\bar x - \bar y$.
For $d\to\io$, the interaction term is dominated by a saddle point on $\bar u$ and $\bar v$, such that
$(x_a - x_b)^2 = (y_a - y_b)^2 = \D_{ab}$ and $x_a \cdot y_b =0$~\cite{KPZ12,KPUZ13,CKPUZ13}, hence
$(x_a - y_a -x_b + y_b)^2 = (x_a - x_b)^2 + (y_a - y_b)^2 = 2 \D_{ab}$. We therefore
obtain at the saddle point
\beq\label{FFreplica}
\begin{split}
\mathcal F_{\{\g_a\}}(\hat\D) 
&=\lim_{n\to 0}\sum_{n_1,\ldots, n_m; \sum_a^mn_a=n}\frac{n!}{n_1!\ldots n_m!}
e^{- \frac{1}{2} \sum_{a,b}^{1,m}\frac{n_an_b}{n^2} \D_{ab} } 
\int \frac{ d\z }{\sqrt{2\pi}}
e^{ - \frac{\z^2}2 \left[ 1 + \frac12 \sum_{ab} \frac{n_a n_b}{n^2} (\g_a - \g_b)^2 \right]    } \\
&=\int \frac{ d\z }{\sqrt{2\pi}}
e^{ - \frac{\z^2}2 }
\lim_{n\to 0}\sum_{n_1,\ldots, n_m; \sum_a^mn_a=n}\frac{n!}{n_1!\ldots n_m!}
e^{- \frac{1}{2} \sum_{a,b}^{1,m}\frac{n_an_b}{n^2} \left( \D_{ab} + \frac{\z^2}{2} (\g_a - \g_b)^2 \right) } 
 \\
&=\int \frac{ d\z }{\sqrt{2\pi}}
e^{ - \frac{\z^2}2 }
\FF\left( \D_{ab} + \frac{\z^2}{2} (\g_a - \g_b)^2 \right) \ ,
\end{split}\eeq
where $\FF(\hat\D)$ is the interaction function in absence of shear-strain, that was used 
in~\cite{KPZ12,KPUZ13,CKPUZ13}.

\subsection{Replicated entropy in presence of shear-strain}

Collecting all terms, we obtain for the replicated free energy
\beq\label{eq:repent}
s(\hat\D,\{\g_a\}) = -\b F(\hat\D,\{\g_a\}) = s_{\rm ent}(\hat\D) 
- \frac{d \wh\f}{2} \mathcal F_{\{\g_a\}}(\hat\D) 
 = s_{\rm ent}(\hat\D) 
- \frac{d \wh\f}{2} \int \frac{ d\z }{\sqrt{2\pi}}
e^{ - \frac{\z^2}2 }
\FF\left( \D_{ab} + \frac{\z^2}{2} (\g_a - \g_b)^2 \right)
\ ,
\eeq
where the ``entropic'' or ideal gas term $s_{\rm ent}(\hat\D)$ does not depend on shear-strain
and has been computed in~\cite{KPZ12,KPUZ13,CKPUZ13}.

Eq.~\eqref{eq:repent} has two important properties. First, the correction in $\g_a$ is clearly
quadratic, hence according to Eq.~\eqref{eq:Fexp} we have $\s_a=0$ for all $a$. We will show 
below that $\s_a$ is the average stress of glassy states under a small shear-strain $\g_a$. But because
of disorder, taking the average over all glasses, the average stress vanishes at $\g_a=0$,
hence $\s_a=0$. Second, a uniform shear-strain $\g_a=\g$ has no effect on the replicated system, which
is correct because the molecules are always in the liquid phase~\cite{MP09,Yo12}.

The fact that the correction is quadratic has another important consequence: to obtain the shear modulus
matrix, we can formally consider $(\g_a - \g_b)^2$ as a small parameter and take the first order correction
in this parameter. This is important, because the matrix $\hat\D$ has to be determined by solving the 
variational equations $\frac{\partial s}{\partial \D_{ab}}=0$ and therefore it also depends on $\{\g_\a\}$. However, at
linear order, the dependence of $\hat\D$ on $\{\g_\a\}$ is irrelevant to compute the shear modulus matrix,
precisely because $\frac{\partial s}{\partial \D_{ab}}=0$ identically\footnote{
This argument can be illustrated as follows. Suppose that $s(\hat\D,\e)$ depends on a single parameter
$\e$. 
The matrix $\hat\D_\e$ is the solution of the equation
\beq\nonumber
\frac{\partial s}{\partial \D_{ab}} =0 \ .
\eeq
Hence, the linear variation of the entropy with respect to $\e$ is 
\beq\nonumber
\frac{d s}{d \e}(\hat\D_\e,\e) = \frac{\partial s}{\partial \e}(\hat\D_\e,\e)
+ \sum_{ab} \frac{\partial s}{\partial \D_{ab}}(\hat\D_\e,\e) \frac{d \D_{ab,\e}}{d\e}
= \frac{\partial s}{\partial \e}(\hat\D_\e,\e)
\eeq
precisely because of the variational condition. Hence in the limit $\e\to 0$ one has
\beq\nonumber
\lim_{\e\to 0}
\frac{d s}{d \e}(\hat\D_\e,\e) = \frac{\partial s}{\partial \e}(\hat\D_{\e=0},\e=0) \ ,
\eeq
and the dependence of $\hat\D_\e$ on $\e$ can be ignored, one only has to compute the partial
derivative with respect to $\e$ and evaluate the result using the unperturbed $\hat\D$.
}.

\subsection{The shear modulus matrix}

Eq.~\eqref{eq:repent} is an exact result for $d\to\io$ and it allows one to compute
all the terms in the expansion for small $\g_a$, that contain the linear and non-linear
responses to an applied shear-strain.
In this paper, however, we focus on the linear response defined by the matrix
$\mu_{ab}$ introduced in Eq.~\eqref{eq:Fexp}.

Because the matrix $\hat\D$ is symmetric, 
we consider that $\FF(\hat\D)$ is a function of $\D_{a<b}$ only,
and the derivative $\frac{\partial \FF}{\partial \D_{ab}}$ is defined
as the variation of $\FF(\hat\D)$ under a variation of $\D_{ab}$, that therefore
also induces an identical variation of $\D_{ba}$.
With this convention, we have at the lowest order
\beq
-\b F(\hat\D,\{\g_a\}) = -\b F(\hat\D) - \frac{d\wh\f}4 \sum_{a<b} \frac{\partial \FF}{\partial \D_{ab}}
(\g_a - \g_b)^2 \ ,
\eeq
and rearranging the terms it is easy to see that
\beq\label{muab_3}
\b \mu_{ab}  = \frac{d}2 \, \wh\f \left[ \d_{ab} \sum_{c (\neq a)}  \frac{\partial \FF}{\partial \D_{ac}}  - (1-\d_{ab})  \frac{\partial \FF}{\partial \D_{ab}}  \right] \ .
\eeq
Note that from Eq.~\eqref{muab_3} it is clear that $\sum_b \mu_{ab} =0$, as it should be because the molecular liquid is in fact a liquid~\cite{Yo12}.

\subsection{The fullRSB structure}

We can now insert in Eq.~\eqref{muab_3} the fullRSB structure of the matrix $\hat\D$, 
which is the most general structure that appears in the exact solution of the problem in $d\to\io$,
as described in~\cite{CKPUZ13}.
Let us define the matrix $I^{m_i}_{ab}$, which has elements equal to 1 in blocks of size $m_i$ around the diagonal, and zero otherwise.
Then, using the notations of~\cite{CKPUZ13}, and noting that $I^{m_k=1}_{ab} = \d_{ab}$, one has for a $k$RSB matrix:
\beq\label{eq:DabkRSB}
\D_{ab} = \sum_{i=1}^k \wh\D_i  (  I^{m_{i-1}}_{ab} - I^{m_i}_{ab}     )   =  \sum_{i=0}^{k-1} (\wh \D_{i+1} - \wh \D_{i} ) I^{m_{i}}_{ab}  - \wh\D_k \d_{ab} \ .
\eeq
Correspondingly, one has
\beq
\frac{\partial \FF}{\partial \D_{ab}} = \sum_{i=1}^k \frac{2}{m (m_{i-1} - m_i)} (  I^{m_{i-1}}_{ab} - I^{m_i}_{ab}     ) \frac{\partial \FF}{\partial \wh\D_i} \ ,
\eeq
and therefore, introducing $J_{ab}^{m_i} = I_{ab}^{m_i} - m_i \d_{ab}$, we have 
\beq
\b \mu_{ab}  = d \, \wh\f \left[
\sum_{i=1}^k \frac{1}{m (m_{i-1} - m_i)} (  J^{m_{i}}_{ab} - J^{m_{i-1}}_{ab}     ) \frac{\partial \FF}{\partial \wh\D_i} 
\right] \ .
\eeq
Using the results in~\cite[Eqs.~(45) and (61)]{CKPUZ13}, we have
\beq
 \frac{\partial \FF}{\partial \wh\D_i} = \int_{-\io}^\io dh e^h \frac{d}{dh} \frac{\partial g(m,h)}{\partial \wh\D_i} = 
\frac{m(m_i - m_{i-1})}2  \int_{-\io}^\io dh e^h \frac{d}{dh} N_i(m,h) \ ,
\eeq
and following the same procedure as in~\cite[Eqs.~(66)-(70)]{CKPUZ13}, we obtain
\beq
 \frac{\partial \FF}{\partial \wh\D_i} =
\frac{m(m_{i-1} - m_{i})}2 e^{-\wh\D_1/2} \int_{-\io}^\io dh P(m_i,h) f'(m_i,h)^2 \ ,
\eeq
and
\beq
\b \mu_{ab}  = \frac{d}2 \, \wh\f
\sum_{i=1}^k  (  J^{m_{i}}_{ab} - J^{m_{i-1}}_{ab}     )  e^{-\wh\D_1/2} \int_{-\io}^\io dh P(m_i,h) f'(m_i,h)^2
\ .
\eeq
Using then the changes of notations of~\cite[Sec.~VI A]{CKPUZ13}, we obtain
\beq
\b \mu_{ab}  = \frac{d}2 \, \wh\f
\sum_{i=1}^k  (  J^{m_{i}}_{ab} - J^{m_{i-1}}_{ab}     ) \frac1{m^2}  \int_{-\io}^\io dh e^h \wh P(y_i,h) \wh f'(y_i,h)^2
= d
\sum_{i=1}^k  (  J^{m_{i}}_{ab} - J^{m_{i-1}}_{ab}     ) \frac{\wh\kappa_i}{m^2}  \ .
\eeq
Recalling that $m_k=1$, $I_{ab}^{m_k} = \d_{ab}$, and using~\cite[Eq.~(113)]{CKPUZ13},
we can rearrange the expression above as
\beq\label{eq:muab_fin}
\begin{split}
\b \wh\mu_{ab} = \frac{\b \mu_{ab}}d  & = \frac{1}{m^2} \left[
\wh\k_k + \sum_{i=1}^k (m_{i-1} - m_i) \wh\k_i
\right] I_{ab}^{m_k}
+ 
\sum_{i=1}^{k-1} \frac{\wh\k_i - \wh\k_{i+1}}{m^2} I_{ab}^{m_i} - \frac{\wh\k_1}{m^2} I_{ab}^{m_0} \\
& = \frac{1}{m \wh\g_k } I_{ab}^{m_k}
+ 
\sum_{i=1}^{k-1} \frac{1}{m  m_i} \left( \frac1{\wh\g_i} - \frac1{\wh\g_{i+1}}   \right) I_{ab}^{m_i} - \frac{1}{m^2 \wh\g_1} I_{ab}^{m_0} \ .
\end{split}\eeq

\section{Physical interpretation: decomposition of the Gibbs measure over pure states}

We now derive a physical interpretation of Eq.~\eqref{eq:muab_fin} by means of the decomposition of the equilibrium Gibbs measure
over the set of pure states, each representing a glassy state. Since this decomposition has been discussed in detail in the 
literature~\cite{MPV87,MP09,PZ10,Yo12},
we assume that the reader is familiar with the construction.
For simplicity, we discuss here the 2RSB case only, by generalizing the 1RSB discussion of~\cite{Yo12}:
the generalization to $k$RSB and fullRSB is straightforward.

\subsection{Decomposition of thermodynamic averages over pure glassy states}

\begin{figure}
\includegraphics[width=0.6\textwidth]{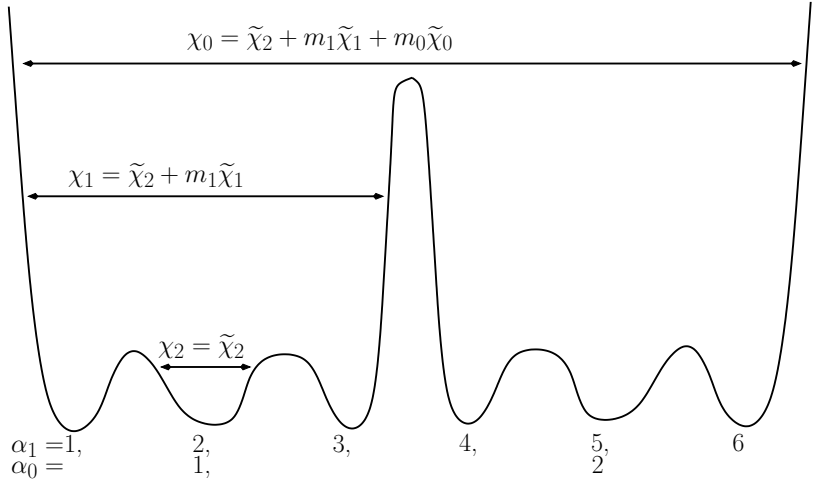}
\caption{
An illustration of the susceptibilities associated to different levels of exploration of phase space in a 2RSB structure
(here with 2 metabasins each split in 3 sub-basins).
A system confined in a single glass basin is characterized by the intra-state susceptibility $\chi_2 = \wt\chi_2$. A system that can explore
the first level of metabasins is characterized by $\chi_1 = \wt\chi_2 + m_1\wt\chi_1$. Finally, a system that can explore the full equilibrium structure
of all metabasins has the equilibrium susceptibility $\chi= \chi_0 = \wt\chi_2 + m_1 \wt\chi_1 + m_0 \wt\chi_0$.
}
\label{fig:basins}
\end{figure}

The 2RSB ansatz describes a system whose glassy states are arranged as follows (see Fig.~\ref{fig:basins}). 
Each glass state (or basin) is labeled by an index $\a_1$ and
has a free energy per particle $f_{\a_1}$. These states are organized in ``metabasins''. Each metabasin is labeled by an index $\a_0$ and contains
a set of glassy basins. We say that $\a_1 \in \a_0$ if basin $\a_1$ belongs to metabasin $\a_0$.
A basin $\a_1$ has a weight within its metabasin $\a_0$ given by
\beq
w_{\a_1 | \a_0} = \frac{e^{-\b N m_1 f_{\a_1} }}{ \sum_{\a_1 \in \a_0} e^{-\b N m_1 f_{\a_1} }} = \frac{e^{-\b N m_1 f_{\a_1} }}{  e^{-\b N m_1 f_{\a_0} }} \ ,
\eeq
where $T/m_1$ is the effective temperature associated to the distribution of basins within metabasins, and
we defined
\beq
Z_{\a_0} = e^{-\b N m_1 f_{\a_0} } =  \sum_{\a_1 \in \a_0} e^{-\b N m_1 f_{\a_1} }
\eeq
as the partition function restricted to metabasin $\a_0$. 
The weight of a metabasin in the total Gibbs measure is given by
\beq
w_{\a_0} = \frac{e^{-\b N m_0 f_{\a_0} }}{\sum_{\a_0} e^{-\b N m_0 f_{\a_0} }} =
\frac{\left( Z_{\a_0}\right)^{\frac{m_0}{m_1}} }{Z} \ ,
\eeq
where $T/m_0$ is the effective temperature associated to the distribution of metabasins, and
the total partition function of the system is 
\beq\label{eq:Z}
Z = \sum_{\a_0} e^{-\b N m_0 f_{\a_0}} = \sum_{\a_0} \left( Z_{\a_0} \right)^{\frac{m_0}{m_1}} = \sum_{\a_0} \left(  \sum_{\a_1 \in \a_0} e^{-\b N m_1 f_{\a_1} } \right)^{\frac{m_0}{m_1}} \ .
\eeq
Finally, the free energy of the system is
\beq
F = - \frac{T}{N m_0} \log Z \ ,
\eeq
where we multiply by $T/m_0$ which is the effective temperature of metabasins.

We now suppose that a field $h$, conjugated to an observable $O$, 
is added to the Hamiltonian of the system which becomes $H' = H - h O$.
One can show by a simple calculation that
\beq\label{eq:dFdh}
\langle O \rangle = - \frac{dF}{dh} =-  \sum_{\a_0} w_{\a_0} \sum_{\a_1\in\a_0} w_{\a_1 | \a_0} \frac{df_{\a_1}}{dh} =  \langle \langle O_{\a_1} \rangle_1 \rangle_0 \ ,
\eeq
where we defined $O_{\a_1} = - \frac{df_{\a_1}}{dh}$ the average of the observable $O$ within the glassy state $\a_1$,
$\langle \bullet \rangle_1$ the average over basins in a metabasin with weights $w_{\a_1}$, $\langle \bullet \rangle_0$ the average
over metabasins with weights $w_{\a_0}$, and $\langle \bullet \rangle$ the total equilibrium average.
The interpretation of Eq.~\eqref{eq:dFdh} is straightforward: the total thermodynamic average of observable $O$ in the Gibbs measure is obtained
by first taking the average within each glassy basin, then taking the average over basins in a metabasin, and finally by taking the average over metabasins.

Taking another derivative, and defining as $\chi_{\a_1} = \frac{d O_{\a_1}}{dh} = - \frac{d^2 f_{\a_1}}{dh^2}$ 
the susceptibility inside state $\a_1$,
we obtain
\beq\label{eq:chiGibbs}
\begin{split}
\chi = \frac{d\langle O\rangle}{dh} =
- \frac{d^2 F}{dh^2} =& 
\sum_{\a_0} w_{\a_0} \sum_{\a_1\in\a_0} w_{\a_1 | \a_0} \frac{d O_{\a_1}}{dh} 
+\sum_{\a_0} w_{\a_0} \sum_{\a_1\in\a_0} \frac{d w_{\a_1 | \a_0}}{dh} O_{\a_1}
+\sum_{\a_0} \frac{d w_{\a_0}}{dh}\sum_{\a_1\in\a_0} w_{\a_1 | \a_0} O_{\a_1} \\
& =  \langle \langle \chi_{\a_1} \rangle_1 \rangle_0 +
\b N m_1
\langle \,   \langle O_{\a_1}^2 \rangle_1 -  \langle O_{\a_1} \rangle_1^2  \, \rangle_0
+\b N m_0 \left[
\langle \langle O_{\a_1} \rangle_1^2 \rangle_0 - \langle \langle O_{\a_1} \rangle_1 \rangle_0^2
\right] \\ 
& = \wt\chi_2 + m_1 \wt\chi_1 + m_0 \wt\chi_0 \ ,
\end{split}\eeq
where
\beq\label{eq:chi012}
\begin{split}
\wt\chi_2 & = \langle \langle \chi_{\a_1} \rangle_1 \rangle_0 =  \sum_{\a_0} w_{\a_0} \sum_{\a_1\in\a_0} w_{\a_1 | \a_0} \frac{d O_{\a_1}}{dh}
\ , \\
\wt\chi_1 & =\b N \langle \,   \langle O_{\a_1}^2 \rangle_1 -  \langle O_{\a_1} \rangle_1^2  \, \rangle_0
=  \b N  \sum_{\a_0} w_{\a_0}  \left[ 
\sum_{\a_1\in\a_0} w_{\a_1 | \a_0} \left( O_{\a_1} \right)^2
-  \left( \sum_{\a_1 \in \a_0} w_{\a_1 | \a_0} O_{\a_1} \right)^2  
\right]
\ , \\
\wt\chi_0 & = \b N \left[
\langle \langle O_{\a_1} \rangle_1^2 \rangle_0 - \langle \langle O_{\a_1} \rangle_1 \rangle_0^2
\right] =
 \b N \left[
\sum_{\a_0} w_{\a_0} \left( \sum_{\a_1 \in \a_0} w_{\a_1|\a_0} O_{\a_1} \right)^2
- \left( \sum_{\a_0} w_{\a_0} \sum_{\a_1\in\a_0} w_{\a_1|\a_0} O_{\a_1}  \right)^2
\right]
\ . \\
\end{split}\eeq
The interpretation of Eqs.~\eqref{eq:chiGibbs} and \eqref{eq:chi012} is also straightforward, and is illustrated in Fig.~\ref{fig:basins}.
The susceptibility $\wt\chi_2$ is the average of the susceptibility of a given glass state $\chi_{\a_1}$ over states: it represents the average susceptibility
of basins.
The term $\wt\chi_1$ is given by the fluctuation of $O_{\a_1}$ from basin to basin inside a same metabasin (averaged over metabasins).
Finally, the term $\wt\chi_0$ is given by the fluctuation from metabasin to metabasin of the average of $O$ within a metabasin, given by $\langle O_{\a_1} \rangle_1$.
An equilibrium system is able to explore all the structure of basins and metabasins, therefore its susceptibility is given by the (properly weighted) sum of 
these three terms. Instead, a system that is completely stuck within a given glass basin will only receive contributions from the term $\wt\chi_2$. A system
that can explore one metabasin, but cannot escape from it, will have a susceptibility $\wt\chi_2 + m_1 \wt\chi_1$.

This construction is obviously generalized to $k$RSB. Recalling that $m_k=1$ corresponds to the innermost basins, while levels $i<k$ corresponds to larger metabasins,
we have that a system that can explore the state structure down to a level $i$ has a susceptibility given by
\beq\label{eq:chii}
\chi_i = \sum_{j=i}^k m_j \wt\chi_j \ .
\eeq
For $i=k$ we have the basin susceptibility, while for $i=0$ we obtain the total susceptibility. Intermediate values of $i$ correspond to partially confined systems.

\subsection{Computing the partial susceptibilities using the molecular liquid}
\label{sec:replicavsstates}

We now discuss the connection with the molecular liquid computation of Sec.~\ref{sec:cloned}.
The molecular liquid provides a way to represent the partition function~\eqref{eq:Z}~\cite{Mo95,MP09}. In a 2RSB ansatz, it is assumed
that replicas arrange in groups of $m_1$ replicas, each group $B =1, \cdots, m_0/m_1$ being in the same state $\a_B$. All replicas are assumed
to be in the same metabasin, hence $\a_B \in \a_0$ for all $B$. Assuming that each replica is subject to a different field $h_a$,
the partition function of the molecular liquid is
\beq
Z_{\rm ml} = \sum_{\a_0}  \prod_{B=1}^{m_0/m_1} \left( \sum_{ \a_1 \in \a_0} e^{-\b N \sum_{a\in B} f_{\a_1}(h_a) } \right) \ ,
\eeq  
and its free energy is
\beq
F_{\rm ml} = -\frac{T}{N} \log Z_{\rm ml}  \ .
\eeq
Note that here we do not divide by $m_0$, which is natural because $Z_{\rm ml}$ is interpreted as the partition function of molecules at temperature $T$.
In the limit $h_a \to 0$, we have
\beq
Z_{\rm ml} = \sum_{\a_0} \prod_{B=1}^{m_0/m_1}  \left( \sum_{ \a_1 \in \a_0}  e^{-\b N m_1 f_{\a_1} } \right) = \sum_{\a_0} \left( \sum_{\a_1 \in \a_0} e^{-\b N m_1 f_{\a_1} } \right)^{\frac{m_0}{m_1}} \ ,
\eeq
which coincides with Eq.~\eqref{eq:Z}.

A simple computation along the lines of the previous section allows one to compute the partial susceptibilities.
We define
\beq\begin{split}
w_{\a_0} &= \frac{1}{Z_{\rm ml}}  \prod_{B=1}^{m_0/m_1} \left( \sum_{ \a_1 \in \a_0} e^{-\b N \sum_{a\in B} f_{\a_1}(h_a) } \right) \ , \\
w^B_{\a_1 | \a_0} &= \frac{e^{-\b N \sum_{a\in B} f_{\a_1}(h_a) }}{\sum_{ \a_1 \in \a_0} e^{-\b N \sum_{a\in B} f_{\a_1}(h_a) }} \ ,
\end{split}\eeq
where $w_{\a_0}$ is the total weight for having all replicas in state $\a_0$, while $w^B_{\a_1|\a_0}$ is the weight for having
the replicas of group $B$ in state $\a_1 \in \a_0$.
Taking a derivative of the free energy with respect to $h_{a}$ we find
\beq
-\frac{d F_{\rm ml}}{dh_{a}}=\sum_{\a_0} w_{\a_0} \sum_{\a_1 \in \a_0} w^{B(a)}_{\a_1 | \a_0} O_{\a_1}(h_a) \ ,
\eeq
where ${B(a)}$ represents the group which contains the replica $a$.
We now take another derivative, and 
observe that $w^{B(a)}_{\a_1 |\a_0}$ depends only on the fields $h_b$ such that $b\in B(a)$,
hence $\frac{dw^{B(a)}_{\a_1 |\a_0}}{dh_b} \propto I^{m_1}_{ab}$.
We find
\beq\begin{split}
 -\frac{d^2 F_{\rm ml}}{d h_a d h_b} &= 
\d_{ab} \sum_{\a_0} w_{\a_0} \sum_{\a_1 \in \a_0} w^{B(a)}_{\a_1 |\a_0} \,  \chi_{\a_1}(h_a)
 + I^{m_1}_{ab}\sum_{\a_0} w_{\a_0} \sum_{\a_1 \in \a_0} \frac{d w^{B(a)}_{\a_1 | \a_0}}{dh_b} O_{\a_1}(h_a)
+\sum_{\a_0} \frac{d w_{\a_0}}{dh_b} \sum_{\a_1 \in \a_0} w^{B(a)}_{\a_1 | \a_0} O_{\a_1}(h_a) \ .
\end{split}\eeq
Developing the derivatives and setting $h_a=0$ it is easy to check that we obtain for the susceptibility matrix:
\beq
\chi_{ab} = -\left. \frac{d^2 F_{\rm ml}}{d h_a d h_b} \right|_{h_a=0}
=\wt\chi_2 \d_{ab} + \wt\chi_1 I^{m_1}_{ab} + \wt\chi_0 I^{m_0}_{ab} \ .
\eeq
Recalling that in the 2RSB case $m_2=1$ and $I^{m_2}_{ab} = \d_{ab}$,
the generalization to $k$RSB is straightforward:
\beq
\chi_{ab} = \sum_{i=0}^k \wt\chi_k I^{m_k}_{ab} \ ,
\eeq
which shows that from the susceptibility matrix of the molecular liquid one can extract the partial susceptibilities $\wt\chi_k$.

\subsection{Shear modulus}

In the case of the shear modulus, from Eq.~\eqref{eq:muab_fin} we deduce that
\beq\label{eq:muabstructure}
\begin{split}
\b \wh\mu_{ab} & = \sum_{i=0}^{k} \b \wt\m_i I_{ab}^{m_i} \ , \\
\b \wt\m_k &= \frac{1}{m \wh\g_k } \ , \\ 
\b \wt\m_i & = - \frac{1}{m  m_i} \left( \frac1{\wh\g_{i+1}} - \frac1{\wh\g_i}   \right) \ , \hskip30pt i=1,\cdots,k-1 \ , \\ 
\b \wt\m_0 & = - \frac{1}{m^2 \wh\g_1}  \ .
\end{split}\eeq
Note that $\wt\m_k$ is positive, as it corresponds to the intra-basin shear modulus, while all the other terms are {\it negative}
(because $\wh\g_i$ is a monotonically decreasing function of increasing $i$~\cite{CKPUZ13}), which is
correct because they represent the {\it softening} of the system due to the fact that it can explore more phase space. 

From this result we can compute the total shear modulus $\wh\mu_i$ 
of a system that is able to explore the structure of metabasins down to
level $i$. We place a hat on
the shear modulus to remember that is has been scaled by $2/d$ as in Eq.~\eqref{eq:muab_fin}.
According to Eq.~\eqref{eq:chii}, we have for $i>0$ :
\beq\label{eq:wtmi}
\b \wh\m_i = \sum_{j=i}^k m_j \b \wt\m_j = \frac{1}{m \wh\g_k } + \sum_{j=i}^{k-1} \frac{1}{m } \left( \frac1{\wh\g_i} - \frac1{\wh\g_{i+1}}   \right)
= \frac{1}{m \wh \g_i} \ ,
\eeq
while, recalling that $m_0=m$,
\beq
\b \wh \m_0 = \b \wh \m_1 - m_0 \frac{1}{m^2 \wh\g_1} = 0 \ ,
\eeq
which is correct because if the system can explore the whole phase space, then it is a liquid and has a zero shear modulus.
Finally, note that in the fullRSB limit $k\to\io$, $\wh\m_i$ has a well defined limit, as it should be for physical quantities (while 
$\wt\m_i$ does not have a finite limit). The index $i$ becomes a continuous index $y$~\cite{CKPUZ13} and we obtain
\beq\label{eq:wtmic}
\b \wh \m(y) = \frac{1}{m \g(y)} \ .
\eeq

\subsection{Summary of the results from the fullRSB replica structure}
\label{sec:rep_summary}

We now summarize the results obtained up to this point using the replica formalism. 
Replicas are used to describe the organization of metastable glassy states sketched in Fig.~\ref{fig:basins} for the 2RSB
case, that becomes a full hierarchical structure in the fullRSB limit~\cite{MPV87}.
The order parameter for the glass transition is the matrix of mean square displacements of replicas, $\D_{ab}$, introduced in
Eq.~\eqref{eq:Dabdef}. In the liquid phase, the replicas can explore all of phase space and $\D_{ab}$ is formally infinite. In the
glass, replicas are confined in the glassy basins and $\D_{ab}$ becomes finite.
In the fullRSB case this matrix has the form~\eqref{eq:DabkRSB} for $k\to\io$, and is therefore parametrized by
a continuous function $\D(y)$~\cite{CKPUZ13}. The index $y \in [1,1/m]$ is the continuous limit of $m_i/m$ and it encodes therefore
the levels of the fullRSB hierarchy of states~\cite{MPV87,CKPUZ13}. The value of the Edwards-Anderson order parameter,
$\D_{\rm EA} \equiv \wh\D_k =  \D(y=1/m)$,
corresponds to the mean square displacement inside each individual glassy state. It must be identified with the long time limit
of the mean square displacement of a system confined within a single glass basin, which we call the ``cage radius'' in the following.
The values of $\D(y)$ for $y < 1/m$ correspond to the mean square displacements between systems confined in different glassy states,
at different levels of the hierarchy~\cite{MPV87}.
The function $\g(y)$ that appeared above is simply related to $\D(y)$, see~\cite[Eq.~(116)]{CKPUZ13}.

As discussed in Sec.~\ref{sec:replicavsstates}, the matrix $\mu_{ab}$ encodes the partial susceptibilities of systems confined to explore a limited portion of
the hierarchy of states. It has the same form of $\D_{ab}$, see Eq.~\eqref{eq:muabstructure}, and the quantities $\wh\mu_i$, which become a continuous function
$\wh\m(y)$ in the fullRSB limit, encode the shear modulus of a system confined to explore the hierarchy of glass basins up to level $y$.
Once again, the value $ \wh \m_{\rm EA} \equiv \wh\m(y=1/m)$ corresponds to the shear modulus of a system confined in a single glass basin,
while values of $\wh\m(y)$ for $y<1/m$ correspond to systems that can explore groups of glassy basins.
Eqs.~\eqref{eq:wtmi} and \eqref{eq:wtmic}, which relate the function $\wh\m(y)$ to the order parameter $\D(y)$, 
are the main technical results of this paper.

\section{Predictions for physical observables}

We now derive the physical consequences of our main result, Eq.~\eqref{eq:wtmic}.
Physically, Eq.~\eqref{eq:wtmic} relates the shear modulus function $\wh\m(y)$ to the mean square displacement function $\D(y)$
as discussed in Sec.~\ref{sec:rep_summary}.
Then, using the 
phase diagram of hard spheres in $d\to\io$ and the results for $\D(y)$ derived in previous works~\cite{KPUZ13,CKPUZ13},
we can immediately derive physical predictions for the shear modulus.

\subsection{The dynamical transition}

As shown in~\cite{KPUZ13,CKPUZ13}, at the dynamical transition, where the glassy states first appear, the system is described by
a 1RSB structure corresponding to a constant $\D(y) = \wh\D_1$. 
Therefore, using~\cite[Eq.(113)]{CKPUZ13}, the shear modulus matrix has the following structure~\cite{YM10,Yo12}:
\beq
\b\wh \mu_{ab} = \frac{1}{m \wh\g_1} \left( \d_{ab} - \frac1m \right) = \frac{1}{\wh\D_1} \left( \d_{ab} - \frac1m \right) \ ,
\eeq
and the intra-state shear modulus is
\beq
\b \wh \m_{\rm EA} \equiv \b \wh\m_1 = \frac{1}{ \wh\D_1 }\equiv \frac{1}{\wh\D_{\rm EA}} \ .
\eeq
We obtain therefore that the shear modulus is proportional to the inverse of the cage radius. This quantity jumps from
$\wh\D_{\rm EA} = \io$ for $\wh\f < \wh\f_{\rm d}$ in the liquid phase, to $\wh\D_{\rm EA} \sim 
\wh\D_{\rm d} - C (\wh\f- \wh\f_{\rm d})^{1/2}$ in the glass phase at $\wh\f>\wh\f_{\rm d}$~\cite{PZ10}. Correspondingly, the shear modulus
has a discontinuous jump at the dynamical transition, and behaves as
\beq\label{eq:mud}
\b \wh\m_{\rm EA}(\wh\f) \sim \b\wh\m_{\rm d} + C'  (\wh\f- \wh\f_{\rm d})^{1/2}
\eeq
just above the glass transition. This result is consistent with the results of Mode-Coupling Theory~\cite{SF11} and previous
computations using replicas~\cite{Yo12,Yo12b,OY13}. 
The presence of a jump is consistent with the numerical results of~\cite{IBS12} and the experimental results of~\cite{Klix12},
although the square root singularity cannot be easily observed numerically or experimentally 
due to the fact that the dynamical transition becomes
a crossover in finite dimensions~\cite{BB11}.

\subsection{Scaling of the intra-state shear modulus around jamming}
\label{sec:jamming}

Around the jamming transition, the system is described by a fullRSB solution with a non-constant $\D(y)$~\cite{CKPUZ13}.
The intra-state shear modulus characterizes
the behavior of a system which is completely confined in a glassy state. It is given, using~\cite[Eq.(113)]{CKPUZ13}, by
\beq\label{eq:muEA}
\b \wh \m_{\rm EA} \equiv \b\wh\m_k = \frac{1}{m \wh\g_k } = \frac{1}{\wh\D_k} \equiv \frac{1}{\wh\D_{\rm EA}} \ ,
\eeq
where $\wh\D_{\rm EA}$ is the cage radius of the glass.
Therefore, we obtain (as in the 1RSB case) the prediction that the shear modulus is proportional to 
the inverse of the cage radius, and to temperature, as it should be for hard spheres where rigidity has an {\it entropic} origin~\cite{IBS12}.
It was shown in~\cite{CKPUZ13} that on approaching the jamming transition, $\wh\D_{\rm EA} \sim p^{-\kappa}$, where $p = \b P/\r$
is the reduced pressure, whose inverse vanishes linearly with density on approaching jamming. We therefore predict that
the shear modulus of the glass diverges on approaching the jamming transition of hard spheres 
(hence for $T\to 0$ and $\wh\f \to \wh\f_j^-$)
as
\beq\label{eq:muEAHS}
\b \wh \m_{\rm EA} \sim p^\kappa \sim | \wh\f - \wh\f_j |^{-\kappa} \ ,
\eeq
with the exponent $\kappa=1.41575$ 
given in~\cite{CKPUZ13}. This result is also nicely consistent with the effective medium approach that has been
developed in~\cite{DLBW14}.

An extension of these results to soft harmonic spheres has been also discussed in~\cite{CKPUZ13}. 
In that case, when temperature
goes to zero, one has $m = T/\t$ with a finite $\tau$, while $\wh\g_k$ remains finite. 
Therefore the shear modulus is finite, 
as it should be because in this case rigidity
has a {\it mechanical} origin~\cite{IBS12}, 
and is given by
\beq
\wh \m_{\rm EA} = \frac{\t}{ \wh\g_k } \ .
\eeq
Furthermore, according to the analysis of~\cite{CKPUZ13},
on approaching jamming from above, $\wh\f \to \wh\f_j^+$,
$\t \sim P \sim |\wh\f -\wh\f_j|$ vanishes linearly with pressure $P$ and distance from jamming.
It was also shown in~\cite{CKPUZ13} that for $\wh\f \to \wh\f_j^+$ one has
$\wh\D_{\rm EA} \propto (T/P) \d z$, where $\d z = z -2d$ is the excess of particle contacts with respect
to the isostatic value $z=2d$.
Using this we obtain
the prediction:
\beq
\wh \m_{\rm EA} \propto \frac{P}{\d z} \ ,
\eeq
which is in agreement with the results of~\cite{Wyart,DLDLW14,DLBW14}.

It has also been shown in~\cite{CKPUZ13} that
$\wh\g_k \sim \d z \sim \t^{\nu(\k-1)}$, but unfortunately the exponent $\nu>0$ was not determined.
We therefore obtain for $\wh\f \to \wh\f_j^+$ that
$\wh \m_{\rm EA} = \t/\wh\g_k  \sim \t^{1 + \nu(1 - \k)} \sim |\wh\f -\wh\f_j|^{1+\nu(1 - \k)}$.
It was already noted in~\cite{CKPUZ13} that the choice $\nu (\k-1) = 1/2$ would reproduce
the results of~\cite{Wyart,DLDLW14,DLBW14}, that $\d z  \sim |\wh\f -\wh\f_j|^{1/2}$
and $\wh \m_{\rm EA} \sim |\wh\f -\wh\f_j|^{1/2}$. Future work will surely address this issue through
a precise determination of $\nu$.

\subsection{Exploring the metabasin: out of equilibrium dynamics}

Another very interesting prediction of Eq.~\eqref{eq:wtmic} is that 
if the system is allowed to explore part of the metabasin structure around a glass state, then the shear modulus changes dramatically. 
In fact, suppose that the system is able to explore the metabasins structure down to scale $y$. 
Then the effective shear modulus is given by Eq.~\eqref{eq:wtmic} with some finite $y$. On approaching jamming, if $y$ remains finite, then $\g(y)$ is also
finite~\cite{CKPUZ13}. At the same time, $m \sim 1/p$~\cite{CKPUZ13}, 
and we conclude that the shear modulus scales with pressure as
\beq
\b \wh \m(y) = \frac{1}{m \g(y)} \sim p \ ,
\eeq
i.e. it scales with pressure with a different exponent than the intra-basin shear modulus. Since $\k>1$ in Eq.~\eqref{eq:muEAHS}, 
the metabasin shear modulus is much smaller
than the intra-basin one: exploring part of the phase space makes the system much softer.

How can the system explore the metabasin structure? If the system is prepared at equilibrium in a glassy state, then the barriers that need to be crossed
to change state within a metabasin diverge with $d$ as $d^\a$ with some exponent $\a<1$, making the exploration of phase space impossible.
This is also the case if the glass is prepared through a slow annealing: in this case at the end of the annealing the system is stuck into a glass and cannot
escape from it. In that case the relevant quantity is $\wh \mu_{\rm EA}$ discussed above.
However, if the system is prepared by a fast quench from the liquid into the glass phase at time $0$, 
and its evolution is recorded as a function of the waiting time $t_w$ 
after the quench, then it can explore a bit the metabasin structure before getting stuck into a glass~\cite{Cuku2}.
Suppose that we let the system age after the quench for a time $t_w$, and at time $t_w$ we add a small shear-strain $\g$
to measure the shear modulus, $\mu(t,t_w) = \s(t,t_w)/\g$, 
where $\s(t,t_w)$ is the stress at time $t$.
The evolution of $\mu(t,t_w)$ is illustrated in Fig.~\ref{fig:master}.
We expect that at short $t-t_w$, $\mu(t,t_w)$ reaches a plateau at $\mu(t,t_w) \sim \wh\mu_{\rm EA} \sim p^{\k}$. 
At larger times $t \sim \t_{\rm mb}(t_w)$, the system can explore
neighboring glass basins~\cite{Cuku2}, and $\mu(t,t_w)$ should drop to a second plateau corresponding to 
$\wh\mu(y)$ for some finite $y$, hence $\mu(t,t_w) \sim \wh\mu(y) \sim p$ at longer times.
The scaling $\mu \sim p$ is captured
by the 1RSB solution~\cite{Yo12b,OY13}.
Note that at {\it much} longer times $t \sim \t_\a(t_w)$, $\mu(t,t_w)$ will eventually drop to zero.

The two time scales $\t_{\rm mb}(t_w)$ and $\t_\a(t_w)$ certainly grow with $t_w$, but their scaling is still unclear.
On a mean field level,
it is reasonable to expect that $\t_{\rm mb}(t_w)$, that is the time needed to cross barriers between basins inside a same metabasin,
will grow as a power law of $t_w$~\cite{Cuku2}.
What happens in non-mean-field systems remains however unclear. In Refs.~\cite{YM12,FM13,FP13,BCTT13,BU14}
it was suggested that in low dimensions, the behavior of glasses might be similar
to a spin glass model in presence of an external field or a random-field Ising model.
If this is the case, 
the dynamics can be extremely slow.
In particular $\tau_{\rm mb}(t_w)$ could naturally grow as a $\log(t_w)$ to some
power instead of a power law.
The time scale $\t_\a(t_w)$ is associated to crossing the largest barriers, those separating the largest metabasins; its scaling with $t_w$ is
not clear and it might be much faster above some critical 
pressure~\cite{Ri13}.
 
A preliminary numerical study of aging close to jamming
has been performed in~\cite{OY13}, and the results look consistent with these predictions. However, a more accurate numerical study should be
 performed to make a precise comparison with the theory.
Note that a scaling of the shear modulus $\mu \sim p$, which could possibly fit in this picture if multiple basins were sampled in the experiment,
has been reported in a previous experiment~\cite{Mason97}. Once again, this point is stimulating but needs further investigation.
 
 \subsection{Reparametrization invariance}

\begin{figure}
\includegraphics[width=0.45\textwidth]{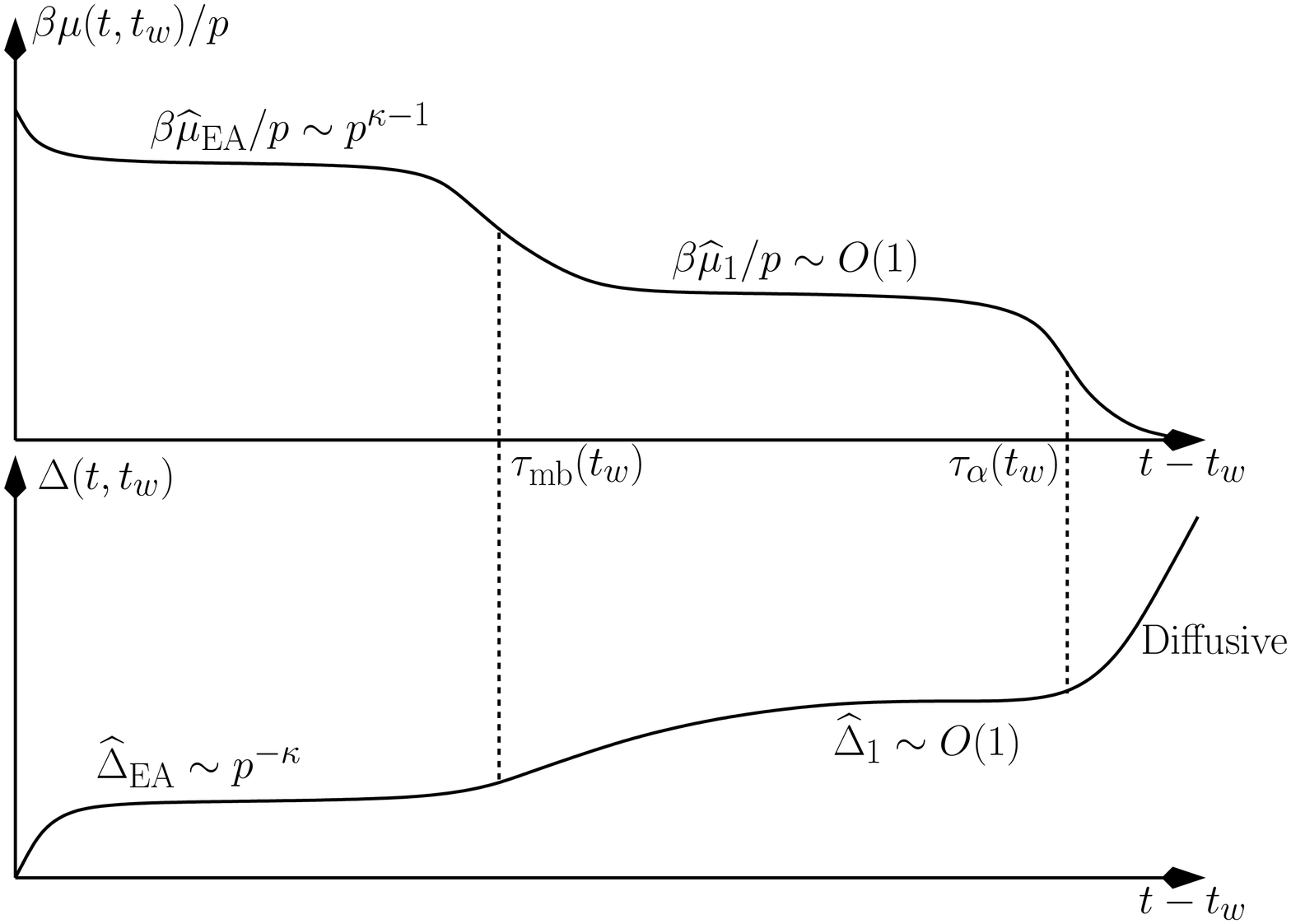}
\includegraphics[width=0.45\textwidth]{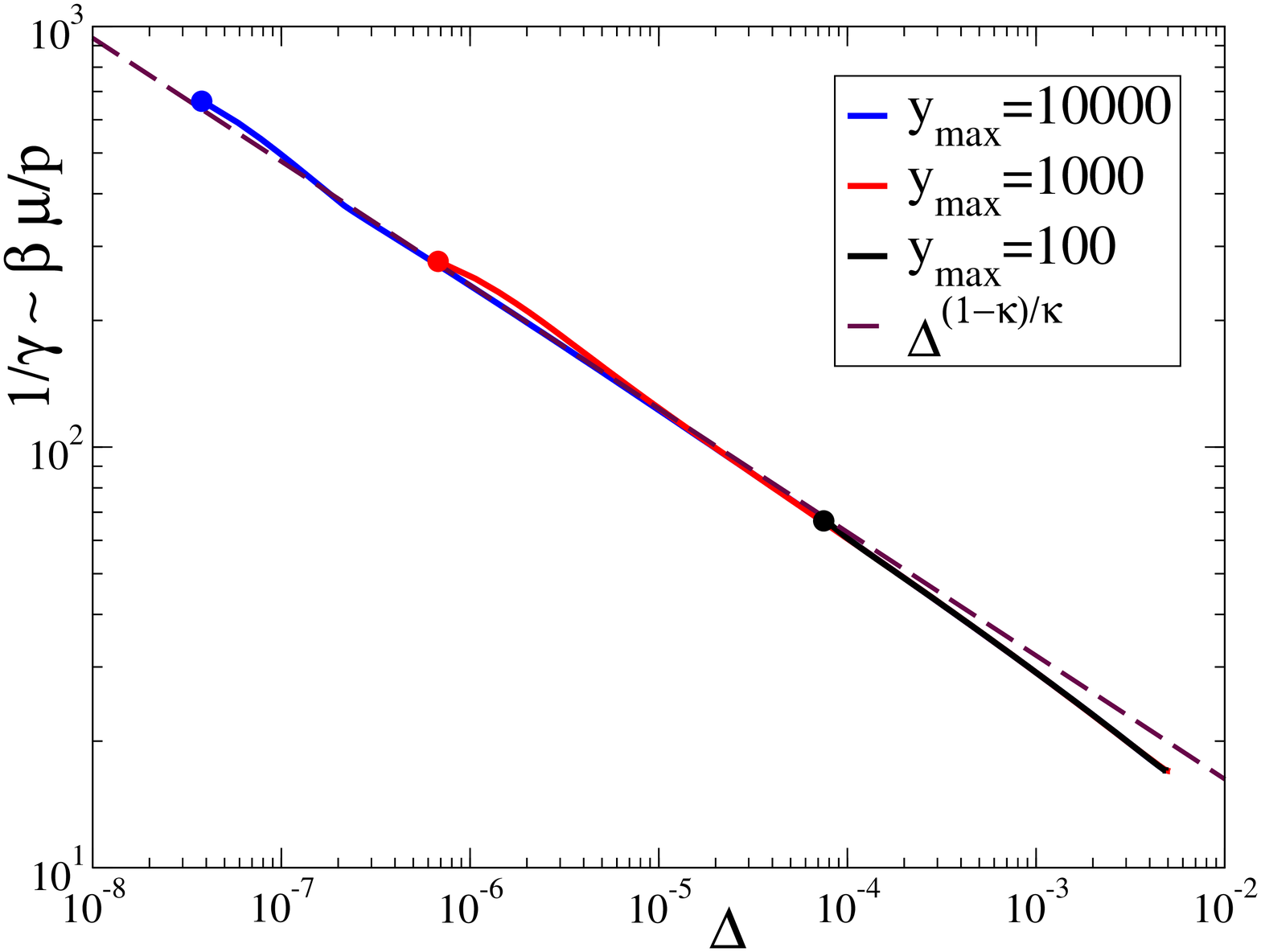}
\caption{
({\it Left}) Schematic evolution of the shear modulus $\mu(t,t_w)$ and the mean square displacement $\D(t,t_w)$,
for a large time $t_w$ after a quench, as a function of $t-t_w$.
({\it Right})
The function $\frac{\b \wh \m(\D)}{p} \sim  1/\g(\D)$ that characterizes the reparametrization invariant regime,
obtained from the numerical solution of the fullRSB equations~\cite{CKPUZ13}. The dots mark the intra-state point
$(\wh\D_{\rm EA},\b\wh\m_{\rm EA}/p)$. The three curves correspond to different cutoffs $y_{\rm max} \propto p$. 
The slight bending of the curves around the dots is an artifact of the numerical computation~\cite{CKPUZ13}.
}
\label{fig:master}
\end{figure}

A more stringent test of the theory can be performed by looking at the so-called {\it reparametrization invariant} regime of the out-of-equilibrium dynamics after
a quench. Suppose that we perform the aging experiment described above, and starting from a time $t_w$ after the quench, we measure both the mean
square displacement $\D(t,t_w) = N^{-1} \sum_i \langle [x_i(t) - x_i(t_w) ]^2 \rangle$ and the shear modulus $\mu(t,t_w)$.
After a short transient, both quantities reach their asymptotic values inside a glass basin, hence $\D(t,t_w) \sim \wh \D_{\rm EA}$ and
$\mu(t,t_w) \sim \wh\mu_{\rm EA}$. 
For large $t_w$, and at larger times $t \sim \t_{\rm mb}(t_w)$, the dynamics enter in a {\it reparametrization invariant} regime~\cite{Cuku2}. 
In this regime, larger and larger groups
of states are explored {\it in a restricted equilibrium}. Hence, $\D(t,t_w) \sim \D(y)$ and $\mu(t,t_w) \sim \wh\mu(y)$ with the same value of $y$ that is dynamically
selected. One can then eliminate time $t$ parametrically from the dynamical data
to obtain a curve $\wh\mu(\D, t_w)$. The same can be done in the replica result by 
eliminating $y$ parametrically to obtain a curve $\wh\mu(\D)$.
Based on general results obtained in the context of spin glasses~\cite{Cuku2}, one expects that
\beq
\lim_{t_w \to\io} \wh\mu(\D, t_w) = \wh\mu(\D) \ ,
\eeq 
provided in the replica calculation the parameter $m$ is selected to be on the {\it threshold line}~\cite{Cuku2, CKPUZ13} 
corresponding to the pressure
at which the dynamics is investigated.

The replica prediction is that
\beq
\frac{\b \wh \m(\D)}{p} \propto \b m \wh \m(\D) = \frac{1}{ \g(\D)} \ ,
\eeq
and because $\D(y) \sim y^{-\kappa}$ and $\g(y) \sim y^{-\kappa+1}$ for $y\to\io$, 
we have that for $\D \to 0$, $\frac{\b \wh \m(\D)}{p} \sim 1/\g(\D) \sim \D^{(1-\k)/\k}$.
In other words, in the jamming limit $p\to\io$, $\b \wh\mu(\D)$ is proportional to reduced pressure, 
in such a way that $\frac{\b \wh \m(\D)}{p}$ tends to a finite limit function. At finite pressure, this master function is defined in an interval 
$\D \in [\wh\D_{\rm EA}, \wh\D_1]$.
For large pressures,
$\wh\D_{\rm EA} \sim p^{-\k}$ goes to zero, while $\wh\D_1$ stays finite.
Correspondingly the shorter time plateau of the shear modulus diverges faster
than $p$, leading to the divergent behavior of the master function, $\frac{\b \wh \m(\D)}{p} \sim \D^{(1-\k)/\k}$. 
An illustrative example, obtained from the numerical data
of~\cite{CKPUZ13} for $\g(y)$ and $\D(y)$, is given in Fig.~\ref{fig:master}.

\section{Conclusions}

By means of the exact solution of amorphous infinite-dimensional hard spheres~\cite{KPZ12,KPUZ13,CKPUZ13}, we are able to compute the shear modulus
in the whole phase diagram. We found that in $d\to\io$ the intra-state shear modulus is simply given by Eq.~\eqref{eq:muEA}, $\wh\mu_{\rm EA} = T/\wh\D_{\rm EA}$ and is 
equal to temperature divided by the cage radius of the glass. This formula predicts that, at the dynamical transition, $\wh\mu_{\rm EA}$
jumps from zero to a finite value, followed by a square-root singularity, according to Eq.~\eqref{eq:mud}. Moreover, it predicts that around the jamming transition
$\wh\mu_{\rm EA}$ has critical scalings, described in Sec.~\ref{sec:jamming}, with critical exponents that are predicted analytically~\cite{CKPUZ13}.

Although our results have been obtained in the limit of $d\to\io$, which is the only case where an exact solution is possible,
they agree
with previous analytical results obtained using MCT~\cite{FC02,SF11}, replicas~\cite{Yo12,Yo12b} and effective medium~\cite{DLDLW14,DLBW14} approaches.
We are able to unify these different approaches and put them on a firm theoretical basis, thanks to the fact that the method is exact for $d\to\io$.
Moreover, our predictions are qualitatively consistent with the most recent and detailed numerical investigations~\cite{IBS12}.
Hence, we believe that our results provide a comprehensive picture of the rheology of complex amorphous fluids.

We also analyzed the behavior of the inter-state shear moduli, that characterize the behavior of the system in situations where it is not confined in a single glass
basin, but can also explore neighboring states within larger metabasins. This can happen for example during the out-of-equilibrium dynamics after a sudden
deep quench. An interesting result is that the inter-state shear modulus has a different scaling with pressure on approaching jamming, 
$\b\wh\mu(y) \sim p \ll p^{\k} \sim \b\wh\mu_{\rm EA}$. This means that if the system is able to explore a little bit of phase space beyond a single glass basin,
its rigidity is decreased dramatically. This effect, which is a new prediction, could be detected in numerical simulations and experiments~\cite{OY13}.
We have also discussed the possibility of constructing reparametrization-invariant parametric plots of different observables~\cite{Cuku2}, a procedure
that should allow one to probe the fullRSB structure of the states close to jamming.
Finally, let us recall that the notions of basins and metabasins 
were proposed within the energy landscape picture of structural glasses~\cite{heuer2008exploring}, 
which naturally implies a hierarchy of rigidities~\cite{Yo12}.

There are several points that should be discussed further. 
First of all, this approach can be extended to finite dimensional systems under a mean-field approximation~\cite{MP09,PZ10,Yo12}.
In that case the method is of course approximate, but the qualitative picture (including the scaling properties) is unchanged and the method
provides good quantitative estimates of physical observables, the equation of state of the glass~\cite{MP09,PZ10} and the different transition densities
or temperatures. This will be useful to perform more direct comparisons between the theory and numerical simulations~\cite{IBS12} or experiments~\cite{Mason97,Klix12}.
Most importantly, the exponent $\nu$ that characterises harmonic soft
spheres at zero temperature should be computed,
to check whether the predictions of the present approach are
consistent with an effective medium computation~\cite{Wyart,DLDLW14,DLBW14}.
Of course, most of the conclusions of
Ref.~\cite{CKPUZ13} about the fullRSB solution (e.g. how it is affected by critical fluctuations in finite dimensions) also apply to the present work.

\acknowledgments

The results presented in this work are strongly based on a previous collaboration with
P.~Charbonneau, J.~Kurchan, G.~Parisi and P.~Urbani. We are very grateful for this collaboration and
for many useful discussions
related to this work.
We are very indebted to
Carolina Brito, Eric DeGiuli, Edan Lerner and Matthieu Wyart, for many important exchanges related to their work~\cite{LDW13,DLBW14}, and for sharing with us unpublished
material, that was particularly useful to detect an error in the original version of the manuscript.
We also wish to thank L.~Berthier, G.~Biroli, and A.~Ikeda for many useful discussions and comments.
H. Y. acknowledges financial supports by 
Grant-in-Aid for Scientific Research on Innovative Areas ``Fluctuation \& Structure''
(No. 25103005)  and Grant-in-Aid for Scientific Research (C) (No.  50335337)
from the MEXT Japan, JPS Core-to-Core Program ``Non-equilibrium dynamics of soft matter
and informations''.

\appendix
\section{Fluctuation formula for the shear modulus}
\label{app:A}

\subsection{Fluctuation formula}

In this section
we provide an alternative derivation of Eq.~\eqref{muab_3} based
on the fluctuation formula for the shear modulus matrix
that has been derived in~\cite{Yo12} and reads
\beq
N \mu_{ab} = \sum_{i<j} \langle \mu^a_{ij} \rangle \d_{ab} - 
\b \left(
\sum_{i<j,k<l} \langle \s_{ij}^a \s_{kl}^b \rangle -  \langle \s_{ij}^a \rangle
\langle \s_{kl}^b \rangle
\right) \ ,
\eeq
where indexes $i,j,k,l = 1,\cdots,N$ run over the molecules of the system,
the averages are weighted by the Boltzmann distribution of the molecular liquid and
\beq\begin{split}
\mu_{ij}^a = \mu^a(x_{ai} - x_{aj}) &= 
\left. \frac{\partial^2}{\partial \g_a^2} v(|S(\g_a)(x_{ai}-x_{aj})|) \right|_{\g_a=0} \ , \\
\s_{ij}^a = \s^a(x_{ai} - x_{aj}) &= \left. \frac{\partial}{\partial \g_a} v(|S(\g_a)(x_{ai}-x_{aj})|)\right|_{\g_a=0} \ .
\end{split}\eeq
This formula can be derived by considering the molecular liquid in the canonical ensemble,
without introducing the density field $\r(\olx)$, and taking the second derivative with respect
to shear.

We now introduce a molecular version of the usual $n$-point density functions~\cite{hansen}, which are
defined as
\beq
\r^{(n)}(\olx_1\cdots\olx_n) = \sum_{i_1\neq i_2 \neq \cdots\neq i_n}
\langle \d(\olx_{i_1} - \olx_1) \cdots \d(\olx_{i_n} - \olx_n) \rangle \ ,
\eeq
and give the probability of finding $n$ molecules in positions $\olx_1\cdots \olx_n$. In terms
of these objects, 
and introducing functions $\mu^a(\olx,\oly) = \mu^a(x_{a} - y_{a})$ and
$\s^a(\olx,\oly) = \s^a(x_{a} - y_{a})$,
we have, as in standard liquid theory~\cite{hansen}:
\beq\begin{split}
\sum_{i<j} \langle \mu^a_{ij} \rangle =& \frac12 \int d\olx d\oly \r^{(2)}(\olx,\oly) \mu^a(\olx,\oly) \ , \\
\sum_{i<j} \langle \s^a_{ij} \rangle =& \frac12 \int d\olx d\oly \r^{(2)}(\olx,\oly) \s^a(\olx,\oly) \ , \\
\sum_{i<j,k<l} \langle \s_{ij}^a \s_{kl}^b \rangle  =&
\frac12 \int d\olx d\oly \r^{(2)}(\olx,\oly) \s^a(\olx,\oly) \s^b(\olx,\oly) \\
&+ \int d\olx_1 d\olx_2 d\olx_3 \r^{(3)}(\olx_1,\olx_2,\olx_3) \s^a(\olx_1,\olx_2) \s^b(\olx_1,\olx_3) \\
&+ \frac14  \int d\olx_1 d\olx_2 d\olx_3 d\olx_4 \r^{(4)}(\olx_1,\olx_2,\olx_3,\olx_4) 
\s^a(\olx_1,\olx_2) \s^b(\olx_3,\olx_4) \ .
\end{split}\eeq

We now make use of the fact that, according to the analysis of~\cite{FP99,TS06b} 
in the limit $d\to\io$, many-body correlations factor in products of
two body correlations, and moreover the two-body correlation is given by its first virial contribution.
Equivalently one can say that all $n$-point functions are given by their lowest order virial contribution,
which is~\cite{Sa58}:
\beq
\r^{(n)}(\olx_1\cdots\olx_n) = \prod_{i=1}^n \r(\olx_i) \prod_{i<j}^{1,n} e^{-\b v(\olx_i,\olx_j)} \ ,
\eeq
where
\beq
e^{-\b v(\olx,\oly)} =  \prod_{a=1}^m e^{-\b v(x_a-y_a)}
\eeq
is the replicated interaction potential.

Using this, we get
\beq\label{muab_app0}
\begin{split}
N \mu_{ab} =& 
 \frac12 \int d\olx d\oly \r(\olx)\r(\oly) e^{-\b v(\olx,\oly)} 
[\mu^a(\olx,\oly) \d_{ab} -\b \s^a(\olx,\oly) \s^b(\olx,\oly)] \\
&-\b \int d\olx_1 d\olx_2 d\olx_3 \r(\olx_1)\r(\olx_2)\r(\olx_3) 
e^{-\b v(\olx_1,\olx_2)}e^{-\b v(\olx_2,\olx_3)}e^{-\b v(\olx_3,\olx_1)}
\s^a(\olx_1,\olx_2) \s^b(\olx_1,\olx_3) \\
&-  \frac\b4  \int d\olx_1 d\olx_2 d\olx_3 d\olx_4 \prod_{i=1}^4 \r(\olx_i)
\left[ 
\prod_{i<j}^{1,4} e^{-\b v(\olx_i,\olx_j)}
- e^{-\b v(\olx_1,\olx_2)}e^{-\b v(\olx_3,\olx_4)}
\right]
\s^a(\olx_1,\olx_2) \s^b(\olx_3,\olx_4)
\end{split}\eeq
It is easy to show that the last two lines of the previous equation vanish in $d\to\io$. 
Consider for example the three-body term. The integral is dominated by configurations
where $\olx_1 - \olx_2 \sim D$ is orthogonal to $\olx_1 - \olx_3 \sim D$, 
so that $\olx_1$ and $\olx_3$
are far away and $e^{-\b v(\olx_3,\olx_1)} \sim 1$ and can be neglected.
The remaining integral can be evaluated through a saddle point and to leading order
it is equal to the square of the average of $\s^a$, which vanishes in an isotropic liquid.
Similarly, the last line is dominated by configurations where e.g. 1 and 3 overlap,
1 and 2 (and 3 and 4) are at distance $\sim D$, and 2 and 4 are far away. Again at this leading
order one obtains the square of the average of $\s^a$, which vanishes. This analysis is consistent
with the general principle that all contributions to thermodynamic averages coming from $n$-body
correlation for $n>2$ vanish in high dimensions~\cite{TS06b}. Since the only two body contribution in the
second and third line of Eq.~\eqref{muab_app0} is the average of $\s^a$ which is zero, \
these contributions must vanish.

We obtain
\beq
\mu_{ab} =
 \frac1{2N} \int d\olx d\oly \r(\olx)\r(\oly) e^{-\b v(\olx,\oly)} 
[\mu^a(\olx,\oly) \d_{ab} -\b \s^a(\olx,\oly) \s^b(\olx,\oly)] \ ,
\eeq
and inserting the explicit expressions of $\mu^a$ and $\s^a$ it is easy to check
that this expression becomes
\beq\label{muab_0}
\b \mu_{ab} = -\frac1{2N} \int d\overline{x} d\overline{y} 
\r(\overline{x})\r(\overline{y})
(x_{a2}-y_{a2}) (x_{b2}-y_{b2}) 
\frac{\partial^2 f}{\partial x_{a1}\partial x_{b1}}(\overline{x} - \overline{y}) \ ,
\eeq
where $f$ is the usual replicated Mayer function in absence of shear-strain~\cite{KPZ12},
corresponding to Eq.~\eqref{eq:Mayer} for $\g_a=0$.
We now take into account translational invariance following the discussion of~\cite{KPZ12}.
We introduce coordinates $u_a = x_a - X$ with $X = m^{-1} \sum_a x_a$, and we use that
$\r(\overline{x})$ does not depend on $X$, to obtain, following the notations of~\cite{KPZ12}:
\beq\label{muab_1}
\b \mu_{ab}  = - \frac{1}{2\r} \int dX \DD\overline{u} \DD\overline{v} 
\r(\overline{u})\r(\overline{v})
(X_2 + u_{a2}-v_{a2}) (X_2 + u_{b2}-v_{b2}) 
\frac{\partial^2 f}{\partial u_{a1}\partial u_{b1}}(X + \overline{u} - \overline{v})  \ .
\eeq

\subsection{Simplifications for $d\to\io$}

We now make use of the results of~\cite[Sec.~5]{KPZ12}, 
that show that the integral in Eq.~\eqref{muab_1} is dominated by 
the region where
$u_a \sim v_a \sim d^{-1/2}$, while $X$ is decomposed in a $(d-m)$-component
vector $X_\perp$, orthogonal to the plane defined by $\overline{u} - \overline{v}$,
that is of order $D$, and a $m$-component vector $X_\parallel \sim d^{-1/2}$.
This structure allows us to perform a series of crucial simplifications of Eq.~\eqref{muab_1}. 
It is clear, in fact, that with probability going to 1 for $d\to\io$ with respect
to the random choice of $\overline{u}, \overline{v}$ according to the probability 
density $\r(\overline{u}) \r(\overline{v})$, the direction $\mu=2$ is orthogonal to
all the vectors $u_a$ and $v_a$, and therefore we can neglect the terms $u_{a2}$
and $v_{a2}$ in Eq.~\eqref{muab_1}. This is further supported by the fact that 
the vectors $u_a$ and $v_a$ are small in the limit $d\to\io$ while $X = X_\perp + X_\parallel$ 
remains of order 1 in most directions. We therefore obtain
\beq
\b \mu_{ab} =  - \frac{1}{2\r} \int \DD\overline{u} \DD\overline{v} 
\r(\overline{u})\r(\overline{v})
\frac{\partial^2 }{\partial u_{a1}\partial u_{b1}}
\int dX (X_2)^2 f(X + \overline{u} - \overline{v})
\eeq
We now observe that direction $\mu=2$ is equivalent to any other direction
$\mu>1$, because the shear-strain has been
eliminated and we are now considering an isotropic system. 
Direction $\mu=1$ is still special due to the presence of the derivative,
but we can consider this as a $1/d$ correction, therefore we can write
\beq
\b \mu_{ab} = - \frac{1}{2\r} \int \DD\overline{u} \DD\overline{v} 
\r(\overline{u})\r(\overline{v})
\frac{\partial^2 }{\partial u_{a1}\partial u_{b1}}
\frac{1}d \int dX X^2 f(X + \overline{u} - \overline{v}) \ ,
\eeq
and, recalling once again that $X = X_\perp + X_\parallel$ with 
$X_\perp \sim D$ and $X_\parallel \sim d^{-1/2}$, we have at leading order
\beq
\b \mu_{ab} = - \frac{1}{2\r} \int \DD\overline{u} \DD\overline{v} 
\r(\overline{u})\r(\overline{v})
\frac{\partial^2 }{\partial u_{a1}\partial u_{b1}}
\frac{1}d \int dX X_\perp^2 f(X + \overline{u} - \overline{v}) \ ,
\eeq

Now we specialize to the case of hard spheres, and 
following exactly the same steps of~\cite[Sec.~5, Eqs.(31)-(33)]{KPZ12},
we obtain
\beq
\int dX X_\perp^2 f(X + \overline{u}) = - V_d D^d \, \frac{\int d^{m}X_\parallel \, 
\Theta_{d-m+2}(D^2  - \min_a | X_\parallel + u_a |^2 )}
{\int d^{m}X_\parallel \, \Theta_{d-m}(D^2  - | X_\parallel |^2 )} \ .
\eeq
Next, following the steps of~\cite[Sec.~5.2]{KPZ12}, and introducing $x_a$ by
$u_a = x_a D/\sqrt{d}$,
we obtain
\beq
\int dX X_\perp^2 f(X + \overline{u}) = 
- V_{d} D^{d+2} \ \frac{ \int d^{m}\ee \, e^{  - \frac12 \min_a | \ee + x_a |^2  } }
{ \int d^{m}\ee \, e^{  - \frac12  | \ee |^2  } } =
- V_d D^{d+2} \FF(\overline{x}) =
- V_d D^{d+2} \FF\left( \frac{\sqrt{d}}{D} \, \overline{u} \right) \ ,
\eeq
and
\beq\label{muab_2}
\b \mu_{ab} = \frac{V_d D^{d+2} }{2\r \, d} \int \DD\overline{u} \DD\overline{v} 
\r(\overline{u})\r(\overline{v})
\frac{\partial^2 }{\partial u_{a1}\partial u_{b1}}
 \FF\left( \frac{\sqrt{d}}{D} \, (\overline{u} - \overline{v}) \right) \ .
\eeq

\subsection{Saddle point evaluation}

It has been shown in~\cite{KPZ12} that 
for $d\to\io$, integrals such as \eqref{muab_2} are dominated by a saddle point
on $\overline{u}$ and $\overline{v}$, due to the fact that $\r(\overline{u})$ is exponential
in $d$. Because the function $\FF$ is not exponential in $d$, it does not contribute to the saddle
point. Because $\int \DD\overline{u} \r(\overline{u}) = \r$, we obtain,
recalling that $\r V_d D^d = 2^d \f = d \wh\f$ where $\f$ is the packing fraction and $\wh\f$ a scaled packing
fraction,
\beq
\b \mu_{ab} = \frac{\r V_d D^{d+2} }{2 d}
\frac{\partial^2 }{\partial u_{a1}\partial u_{b1}}
 \FF\left( \frac{\sqrt{d}}{D} \, (\overline{u}^{\rm sp} - \overline{v}^{\rm sp}) \right) 
= \frac{ \wh\f D^{2}}2
\frac{\partial^2 }{\partial u_{a1}\partial u_{b1}}
 \FF\left( \frac{\sqrt{d}}{D} \, (\overline{u}^{\rm sp} - \overline{v}^{\rm sp}) \right) 
\ . 
\eeq
The function $\FF$ is rotationally invariant, hence it depends only on $(u_a - v_a) \cdot (u_b - v_b)$.
We can choose any values of $\overline{u}^{\rm sp}$ and
$\overline{v}^{\rm sp}$, provided they respect the saddle point equations, which state~\cite{KPZ12} that
$u_a \cdot u_b = q_{ab}$, $v_a \cdot v_b = p_{ab}$, and $u_a \cdot v_b = 0$,
hence $(u_a - v_a) \cdot (u_b - v_b) = q_{ab} + p_{ab}$. We have that $\hat q = \hat p$, however when we take the
derivative with respect to $u_a$ we should only differentiate with respect to $\hat q$ and not to $\hat p$. We can simplify this by writing
\beq
\b \mu_{ab} = \frac{\wh\f D^{2}}2
\left. \frac{\partial^2 }{\partial u_{a1}\partial u_{b1}}
 \FF\left( \frac{d}{D^2} \, (\hat q_{ab} + \hat p_{ab} ) \right) \right|_{\hat p = \hat q}
 = \frac{\wh\f D^{2}}4
\frac{\partial^2 }{\partial u_{a1}\partial u_{b1}}
 \FF\left( 2 \frac{d}{D^2} \, \hat q_{ab} \right) 
\ . 
\eeq
Also, following~\cite{KPUZ13,CKPUZ13} we can introduce the matrix 
\beq
\D_{ab} = \frac{d}{D^2} (u_a - u_b)^2 = \frac{d}{D^2} (q_{aa} + q_{bb} - 2 q_{ab}) \ ,
\eeq
and we have, following the definitions of~\cite{CKPUZ13}, that
\beq
\b \mu_{ab}  = \frac{\wh\f D^{2}}4
\frac{\partial^2 }{\partial u_{a1}\partial u_{b1}}
 \FF(  \hat \D ) 
\ ,
\eeq
with $\FF(\hat\D)$ given in~\cite{CKPUZ13}.
We have then
\beq
\frac{\partial }{\partial u_{a1}}
 \FF(  \hat \D )  = \frac{d}{D^2} \sum_{c (\neq a)} 2 ( u_{a1} - u_{c1} ) \frac{\partial \FF}{\partial \D_{ac}}  \ ,
\eeq
and 
\beq
\frac{\partial^2 }{\partial u_{a1}\partial u_{b1}}
 \FF(  \hat \D )  = \frac{d}{D^2} \times
 \begin{cases}
2 \sum_{c (\neq a)}  \frac{\partial \FF}{\partial \D_{ac}} + \sum_{c (\neq a), d (\neq a)} 
\frac{\partial^2 \FF }{\partial \D_{ac}\partial \D_{ad}} 4  \frac{d}{D^2}  ( u_{a1} - u_{c1} )  ( u_{a1} - u_{d1} ) & \text{for } a=b \\
- 2  \frac{\partial \FF}{\partial \D_{ab}} + \sum_{c (\neq a), d (\neq b)} 
\frac{\partial^2 \FF }{\partial \D_{ac}\partial \D_{bd}} 4  \frac{d}{D^2}  ( u_{a1} - u_{c1} )  ( u_{b1} - u_{d1} ) & \text{for } a \neq b \\
\end{cases}
\eeq
In each line of the previous equation, the second term is a factor $1/d$ smaller than the first, because it contains terms like
$\frac{d}{D^2} u_{a1}^2 = \frac{d}{D^2} q_{aa}/d \propto \D /d$, hence it can be neglected. We obtain
\beq
\frac{\partial^2 }{\partial u_{a1}\partial u_{b1}}
 \FF(  \hat \D ) = \frac{2d}{D^2} \left[ \d_{ab} \sum_{c (\neq a)}  \frac{\partial \FF}{\partial \D_{ac}}  - (1-\d_{ab})  \frac{\partial \FF}{\partial \D_{ab}}  \right] \ ,
\eeq
and using this we obtain our final result
\beq
\b \mu_{ab}  = \frac{d}2 \, \wh\f \left[ \d_{ab} \sum_{c (\neq a)}  \frac{\partial \FF}{\partial \D_{ac}}  - (1-\d_{ab})  \frac{\partial \FF}{\partial \D_{ab}}  \right] \ ,
\eeq
which coincides with Eq.~\eqref{muab_3}.

\bibliographystyle{mioaps}
\bibliography{HS}

\end{document}